%% file: ms.tex
\documentclass[useAMS,usenatbib,usegraphicx]{mn2e}

\input{definitions}

\usepackage{url}
\usepackage[varg]{txfonts}
\usepackage{color}
\usepackage{subfigure}

\title[The non-evolving internal structure of early-type galaxies:
  J0728]{The non-evolving internal structure of early-type galaxies:
  the case study SDSS\,J0728\boldmath{$+$}3835 at \boldmath{$z =
    0.206$}}

\author[M. Barnab\`e et al.]{%
  Matteo Barnab\`e$^{1,2}$\thanks{E-mail: mbarnabe@stanford.edu}, 
  Matthew W. Auger$^{2}$,
  Tommaso Treu$^{2,7}$,
  L\'eon V. E. Koopmans$^{3}$,
  \newauthor
  Adam S. Bolton$^{4,5}$, 
  Oliver Czoske$^{3}$ and
  Rapha\"el Gavazzi$^{6}$\\
  \\
  $^{1}$Kavli Institute for Particle Astrophysics and Cosmology,
  Stanford University, 452 Lomita Mall, Stanford, CA 94035-4085, USA\\
  $^{2}$Department of Physics, University of California, Santa
  Barbara, CA 93101, USA\\
  $^{3}$Kapteyn Astronomical Institute, University of Groningen, 
  PO Box 800, 9700\,AV Groningen, Netherlands\\
  $^{4}$Department of Physics and Astronomy, University of Utah, 115
  South 1400 East, Salt Lake City, UT 84112, USA\\
  $^{5}$Beatrice Watson Parrent Fellow, Institute for Astronomy,
  University of Hawai`i, 2680 Woodlawn Dr., Honolulu, HI 96822\\
  $^{6}$Institut d'Astrophysique de Paris, CNRS, UMR 7095,
  Universit\'e Pierre et Marie Curie, 98bis Bd Arago, 75014 Paris,
  France \\
  $^{7}$Packard Research Fellow
}

\begin{document}

\date{Accepted 2010 April 20. Received 2010 April 16; in original form 2010 February 03.}

\maketitle

\label{firstpage}

\begin{abstract}
We study the internal dynamical structure of the early-type lens
galaxy SDSS\,J0728$+$3835 at $z = 0.206$. The analysis is based on
two-dimensional kinematic maps extending out to 1.7 effective radii
obtained from Keck spectroscopy, on lensing geometry and on stellar
mass estimates obtained from multiband \textit{Hubble Space Telescope}
imaging. The data are modelled under the assumptions of axial symmetry
supported by a two-integral distribution function (DF), by applying
the combined gravitational lensing and stellar dynamics code
{\cauldron}, and yielding high-quality constraints for an early-type
galaxy at cosmological redshifts. Modelling the total density profile
as a power-law of the form $\rho_{\mathrm{tot}} \propto 1/r^{\slope}$,
we find that it is nearly isothermal (logarithmic slope $\slope =
2.08^{+0.04}_{-0.02}$), and quite flattened (axial ratio $q =
0.60^{+0.08}_{-0.03}$). The galaxy is mildly anisotropic ($\delta =
0.08\pm0.02$) and shows a fair amount of rotational support, in
particular towards the outer regions.  We determine a dark matter
fraction lower limit of 28~per cent within the effective radius. The
stellar contribution to the total mass distribution is close to
maximal for a Chabrier initial mass function (IMF), whereas for a
Salpeter IMF the stellar mass exceeds the total mass within the galaxy
inner regions. We find that the combination of a NFW dark matter halo
with the maximally rescaled luminous profile provides a remarkably
good fit to the total mass distribution over a broad radial range. Our
results confirm and expand the findings of the SLACS survey for
early-type galaxies of comparable velocity dispersion ($\sigma_{\rm
  SDSS}=214\pm11$ \kms). The internal structure of {\jkeck} is
consistent with that of local early-type galaxies of comparable
velocity dispersion as measured by the SAURON project, suggesting lack
of evolution in the past two billion years.
\end{abstract}



\section{Introduction}
\label{sec:introduction}

Unveiling the mass distribution, dynamical structure and dark matter
content of early-type galaxies is of great interest both as a subject
in its own right, considering their importance in the local Universe,
where a large fraction of the total stellar mass is contained within
E/S0 systems \citep*{Fukugita1998}, and in order to provide stringent
tests for the galaxy formation and evolution models.

It is not surprising, therefore, that nearby early-type galaxies have
been the object of intense study during the last decades, by taking
advantage of the diverse available observational tracers. These
include stellar kinematics (see e.g. \citealt*{Saglia1992},
\citealt{Franx1994}, \citealt{Rix1997},
\citealt{Loewenstein-White1999}, \citealt{Gerhard2001},
\citealt{Borriello2003}, \citealt{Cappellari2007},
\citealt{Thomas2007b}, \citealt{Weijmans2009}), globular clusters and
planetary nebulae kinematics \citep[e.g.][]{Cote2003, Romanowsky2003,
  deLorenzi2008}, the occasional $\ion{H}{I}$ disk or ring
(e.g.\ \citealt*{Franx1994}; \citealt{Weijmans2008}) and hot X-ray
emission \citep[e.g.][]{Matsushita1998, Fukazawa2006, Humphrey2006,
  Humphrey-Buote2010}. The general picture emerging from many of these
studies is that the total mass density profile of ellipticals can be
well described by a power-law form close to $\rho_{\mathrm{tot}}
\propto 1/r^{2}$, often referred to as the isothermal
profile. Moreover, while the inner regions of early-type galaxies are
clearly dominated by the stellar component, the dark matter component
is usually found to play already a non negligible role, with fractions
of approximately $10$ to $40$ per cent of the total mass within an
effective radius. Studies based on stellar population and dynamical
models \citep[e.g.][]{Padmanabhan2004} indicate that the dark matter
fraction increases with the mass of the galaxy, a trend that is more
conspicuous in the case of slow-rotator ellipticals
\citep{Tortora2009}.

The analysis of early-type galaxies beyond the local Universe,
i.e. beyond redshift $z \approx 0.1$, holds great promise in view of
understanding the structural evolution of these objects, but it also
presents several difficulties which hinder the application of
traditional techniques. Stellar dynamics studies, in particular, are
limited by the degeneracy between the galaxy mass profile and the
anisotropy of the stellar velocity dispersion tensor. Taking into
account higher order velocity moments can provide a solution
\citep[see][]{Gerhard1993, vanderMarel-Franx1993, Lokas-Mamon2003},
but unfortunately carrying out such measurement for distant systems is
not viable with the current instruments on 8-10 meter class
telescopes.

However, galaxies at $z \gtrsim 0.1$ have a far greater chance of
acting as strong gravitational lenses \citep*{Turner1984}, thus
providing a very helpful additional diagnostic tool. This is
particularly valuable since it allows an accurate and robust
determination of the total mass enclosed, in projection, within the
region delimited by the Einstein radius
\citep{Kochanek1991}. Unfortunately, the diagnostic power of strong
lensing to constrain internal mass distribution of the deflector is
limited, chiefly by the mass-sheet and mass-slope degeneracies
\citep{Falco1985, Wucknitz2002}, although the latter can be partially
overcome by studying spatially extended lensed sources
\citep[e.g.][]{Warren-Dye2003, Suyu2010}.  A very effective way to
overcome these difficulties and robustly recover various structural
properties of the galaxy is to combine the gravitational lensing
analysis with the complementary constraints provided by stellar
dynamics (see \citealt{Koopmans-Treu2002},
\citeyear{Koopmans-Treu2003}, \citealt{Treu-Koopmans2002b},
\citeyear{Treu-Koopmans2002a}, \citeyear{Treu-Koopmans2004},
\citealt{Barnabe-Koopmans2007}, hereafter BK07, and, e.g.,
\citealt{Rusin-Kochanek2005}, \citealt{Jiang-Kochanek2007},
\citealt{vandeVen2008}, \citealt{Trott2010}, \citealt{Grillo2010} for
further applications of this approach).

Up to very recently, the availability of only a handful of lens
galaxies suitable for the joint analysis represented a major
limitation. This concern has been dispelled by the Sloan Lens ACS
Survey, SLACS (\citealt{Bolton2006}, \citeyear{Bolton2008a},
\citeyear{Bolton2008b}, \citealt{Koopmans2006},
\citeyear{Koopmans2009}, \citealt{Treu2006}, \citeyear{Treu2009},
\citealt{Gavazzi2007}, \citeyear{Gavazzi2008}, \citealt{Auger2009}),
which has led to the discovery of a large and homogeneous sample of
almost a hundred strong gravitational lenses, mostly early-type
galaxies, in the redshift range of $z \approx 0.05 - 0.5$. For a
subset of about~$30$ SLACS systems, the data set is complemented by
two-dimensional kinematic maps of the lens obtained from spectroscopic
observations carried out either with the Very Large Telescope (VLT)
instrument VIMOS or with the Low Resolution Imager and Spectrograph
(LRIS, see \citealt{Oke1995}) mounted on the Keck-I telescope. This
has provided further motivation to expand the combined analysis
technique into a more general and self-consistent method which makes
full use of the available data sets (i.e.\ surface brightness
distribution of both the lensed source and the lens galaxy, and
two-dimensional kinematic maps of the latter), and is coherently
embedded in the framework of Bayesian statistics (BK07).  The current
implementation of the method, the {\cauldron} code --- based on the
assumptions of axial symmetry and two-integral stellar DF for the lens
galaxy --- has been used to conduct an in-depth study of a SLACS
subsample of six systems representative of the survey in terms of
redshifts and velocity dispersions (see \citealt{Czoske2008} and
\citealt{Barnabe2009}, hereafter B09). As shown in those works, the
more sophisticated approach makes it possible to extract much more
information out of the data set, allowing to recover, in addition to
the slope of the total density profile, several other important
properties of the lens galaxies, including the flattening of the
density distribution, lower limits for the dark matter fraction at
different radii and insights on the dynamical structure (angular
momentum, anisotropy, contribution of rotation and random
motions). These quantities are all of relevance to the formation
history of these galaxies.

In this paper we carry out a detailed combined lensing and dynamics
analysis of the SLACS system SDSS\,J0728$+$3835, employing the
{\cauldron} algorithm. The lens is an early-type galaxy at $z =
0.206$, with an aperture averaged velocity dispersion $\sigma = 214
\pm 11$ \kms\ measured from SDSS spectroscopy and a half-light radius
$\Reff = 1.78 \arcsec$ in the $I$~band. The background source is
located at $z = 0.688$ and the Einstein radius is $\REin = 1.25
\arcsec$. With respect to the systems considered in \citet{Czoske2008}
and B09, all followed-up with VLT VIMOS integral-field unit, the main
difference in the observables lies in the kinematic data set: the
{\jkeck} velocity moments maps are obtained from LRIS Keck long-slit
spectroscopic observations, using three slits parallel to the major
axis, and offset along the minor axis, in order to mimic
integral-field capabilities. Remarkably, this kinematic data set
extends significantly farther than those of the previously examined
systems, reaching up to 1.7~$\Reff$, thus providing us important
constraints beyond the inner regions of the galaxy. Moreover, for
the first time, we use the stellar masses determined from stellar
population analysis \citep{Auger2009} to set the normalization of the
luminous mass distribution of the lens galaxy, enabling us to
disentangle the luminous and dark matter contributions and to compare
different choices of the initial mass function (IMF).

The paper is organized as follows: after introducing the data set in
Section~\ref{sec:observations}, we present and discuss the results of
the combined analysis in Section~\ref{sec:analysis} and draw
conclusions in Section~\ref{sec:conclusions}.

Throughout this paper we adopt a concordance $\Lambda$CDM model
described by $\Omega_{\mathrm{M}}=0.3$, $\Omega_{\Lambda} = 0.7$ and
$H_{0} = 100\,h\,\mathrm{km\,s^{-1}\,Mpc^{-1}}$ with $h=0.7$, unless
stated otherwise.


\section{Observations}
\label{sec:observations}

\subsection{High-resolution imaging data}

The lensing analysis requires deep high-resolution imaging data and
this is provided by the SLACS survey's {\it Hubble Space Telescope}
({\it HST}) imaging. In particular, SLACS has obtained one {\it HST}
orbit ($\approx$ 2200s) of data in the F814W filter. The data are
processed as described in \citet{Bolton2008a}; to briefly summarize,
the four individual exposures are background subtracted, cosmic
ray-cleaned, registered, resampled to an output grid with square
pixels that are 0\farcs05 on a side, and stacked with an additional
cosmic ray-rejection step. Synthetic point-spread function (PSF)
images created with TinyTim are likewise resampled and combined to
create a composite model PSF for the output image.

The light distribution of the galaxy is then fit with a B-spline model
\citep[e.g.,][]{Bolton2008a} and this model, convolved with the seeing
present during the spectroscopic observations, is used as the input
surface brightness distribution for the dynamics modelling. The
residual image (the B-spline model subtracted from the data image)
contains the flux from the lensed background source and is used as
lensing constraint on the mass model of the lens galaxy (see BK07 for
details). The galaxy-subtracted image is shown in the bottom
right-hand panel of Fig.~\ref{fig:J0728_LEN}.

\subsection{Pseudo-IFU Spectroscopy}
\label{ssec:IFU}

\begin{figure}
  \begin{center}
    \includegraphics[width=0.45\textwidth,clip]{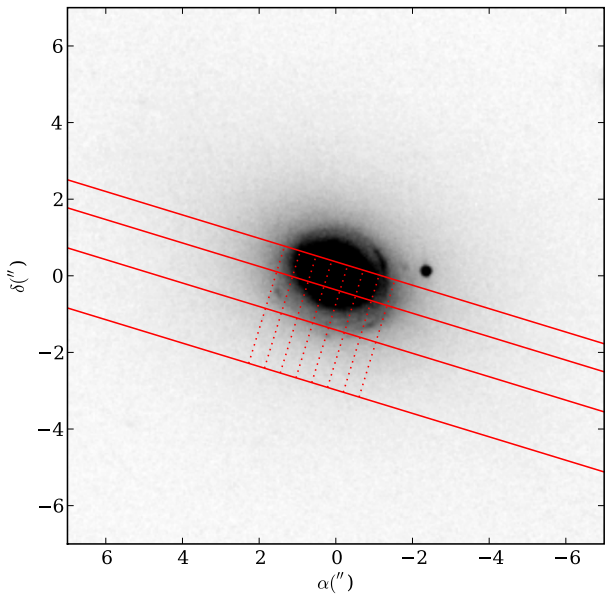}
  \end{center}
  \caption{{\it HST} F814W image of {\jkeck} with the slit locations
    overplotted as solid red lines. The most narrow slit (0\farcs7) is
    centered on the light distribution and follows the semi-major
    axis. The other slits are 1\arcsec and 1\farcs5 wide and are
    offset so as to be adjacent to the previous slit. The dotted
    red lines delineate the 7 apertures with width 0\farcs43 that were
    extracted from each longslit observation.}
  \label{fig:pseudoIFU}
\end{figure}

Our previous joint lensing and dynamics studies used VLT-VIMOS
integral field unit (IFU) spectroscopy to generate resolved stellar
kinematic maps. In this study we employ `pseudo-IFU' spectroscopy,
which consists of dithering a traditional longslit in the spectral
direction over the target galaxy to provide spatially-resolved
kinematic data perpendicular to the slit direction in addition to the
spatial information that is obtained along each slit (see Figure
\ref{fig:pseudoIFU}). This technique yields significantly fewer
spaxels per exposure than proper IFU spectroscopy but the throughput
from the longslit is a factor of a few larger than the IFU throughput
and we are therefore able to efficiently produce kinematic maps,
despite the less efficient way of sampling the kinematic
field. Additionally, we are more sensitive to the low-surface
brightness outer regions of the galaxy and can therefore extend our
kinematic maps to larger radii than is possible with the IFU data.

\begin{figure*}
  \begin{center}
    \includegraphics[width=0.3\textwidth,clip]{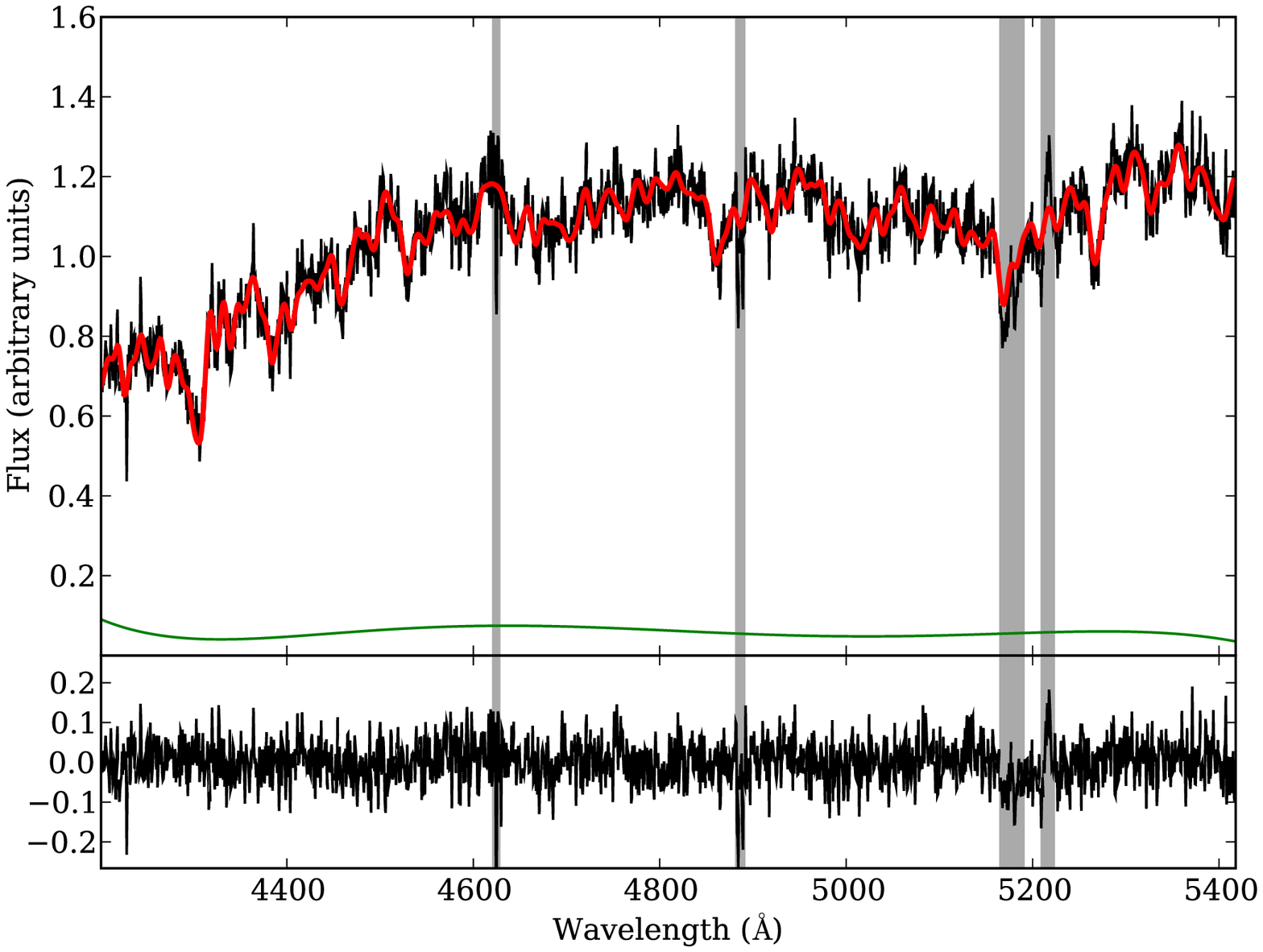}
    \includegraphics[width=0.3\textwidth,clip]{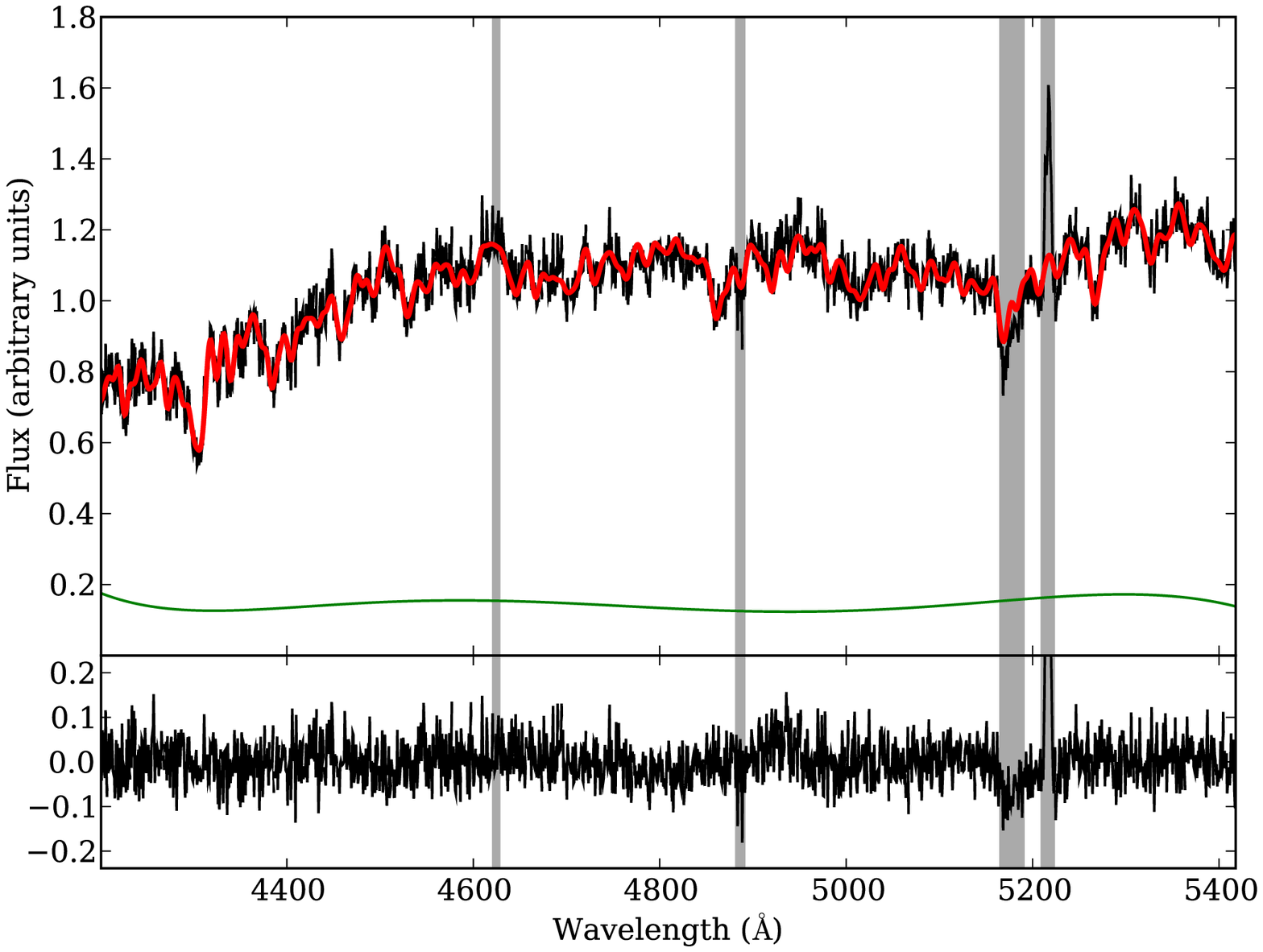}
    \includegraphics[width=0.3\textwidth,clip]{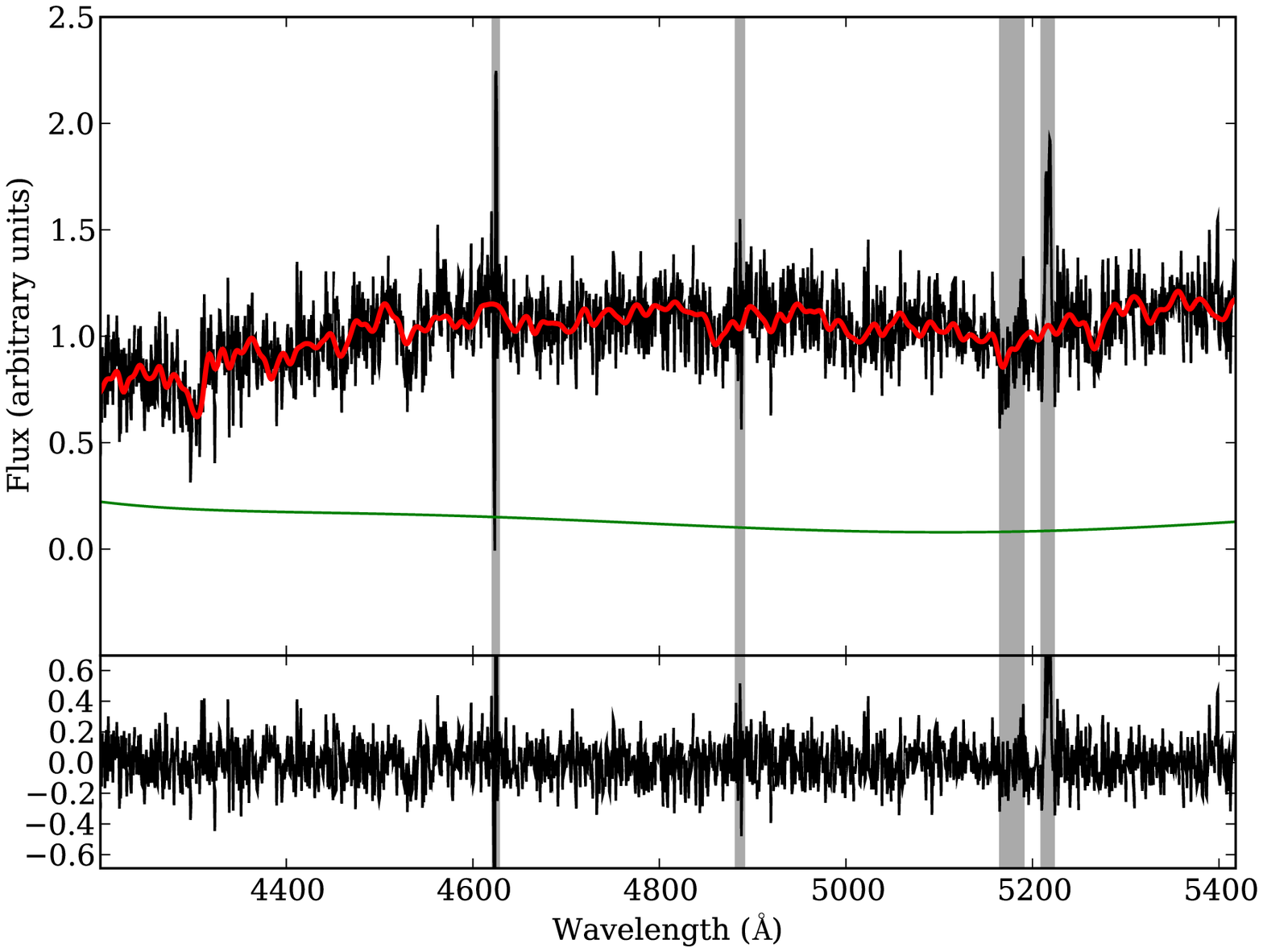}
  \end{center}
  \caption{Extracted one-dimensional spectra for the central aperture
    of the 0\farcs7 slit (left), 1\arcsec slit (center), and 1\farcs5
    slit (right). The red line is the best-fit model, the green line
    is the polynomial continuum model, and the lower panel shows the
    residuals. Greyed-out regions indicate parts of the spectra that
    were not included in the fit (two strong night-sky features, the
    Mgb line of {\jkeck}, and the \ion{O}{II} features from the
    background source).}
  \label{fig:keckSpectra}
\end{figure*}

The spectroscopic data for SDSS J0728 were obtained with LRIS on Keck
I during the nights of 22 and 23 December 2006. The observing
conditions were clear with $\approx 0\farcs8$ seeing, and we used the
460 dichroic to split the beam to the blue and red sides of the
spectrograph. Here we only use data from the red side, which employed
the 900/5500 grating with a dispersion scale of 0.85\,\AA
pixel$^{-1}$. Three slit positions aligned along the semi-major axis
of the galaxy were observed, including: two 900s exposures with a
0\farcs7 slit positioned on the center of the galaxy; four 1200s
exposures with a 1\arcsec slit offset 0\farcs85 from the center of the
galaxy; and 4 1800s exposures with a 1\farcs5 slit offset 2\farcs1
from the center of the galaxy. The data were reduced using custom
Python scripts \citep[for details see][]{Suyu2010} and one dimensional
spectra were extracted from apertures of width 2 pixels ($\approx
0\farcs43$) at seven points along each slit as indicated in Figure
\ref{fig:pseudoIFU}, yielding 21 spectra in total (e.g., Figure
\ref{fig:keckSpectra}).

\begin{figure*}
  \begin{center}
    \includegraphics[width=0.45\textwidth,clip]{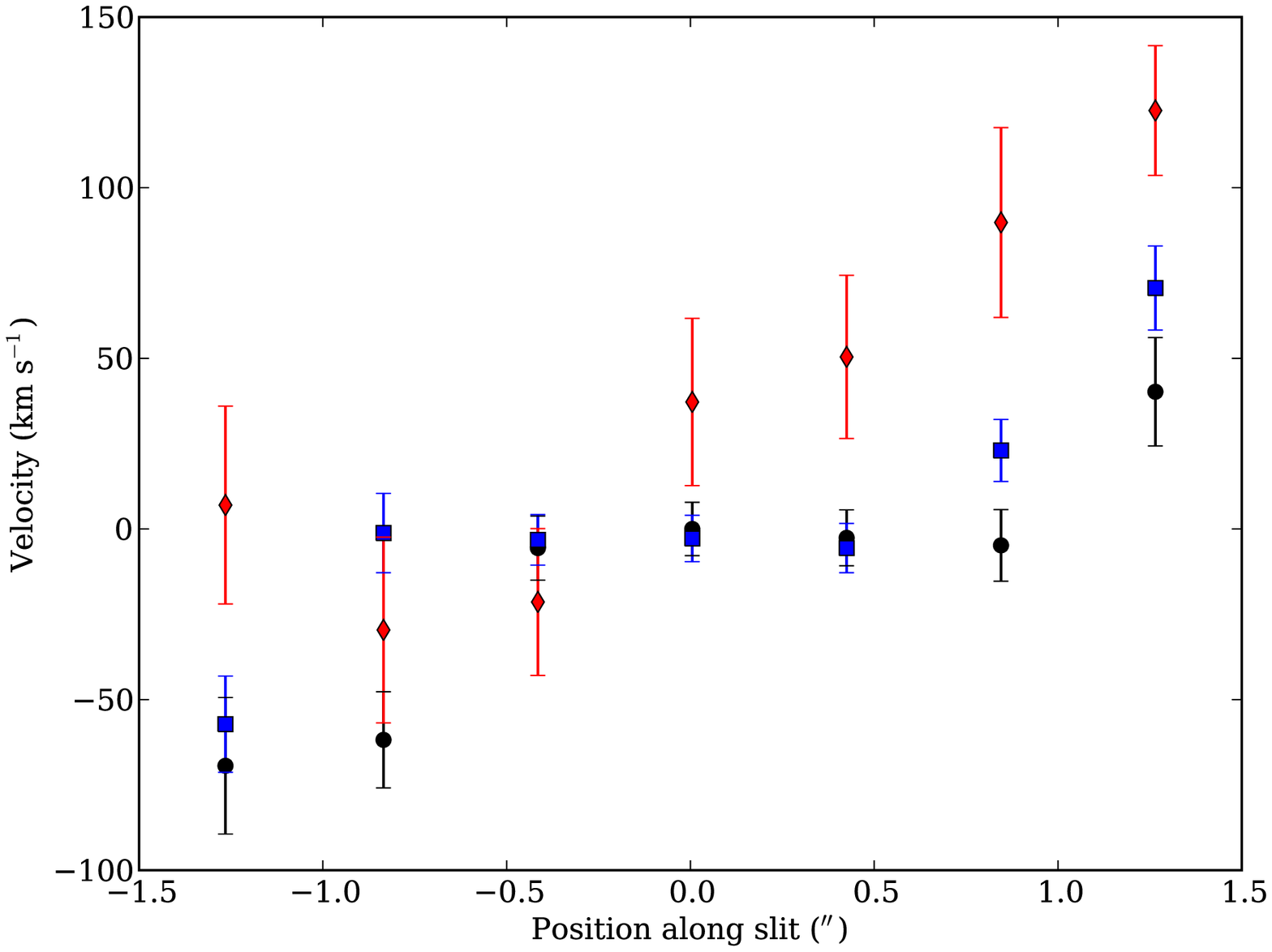}
    \includegraphics[width=0.45\textwidth,clip]{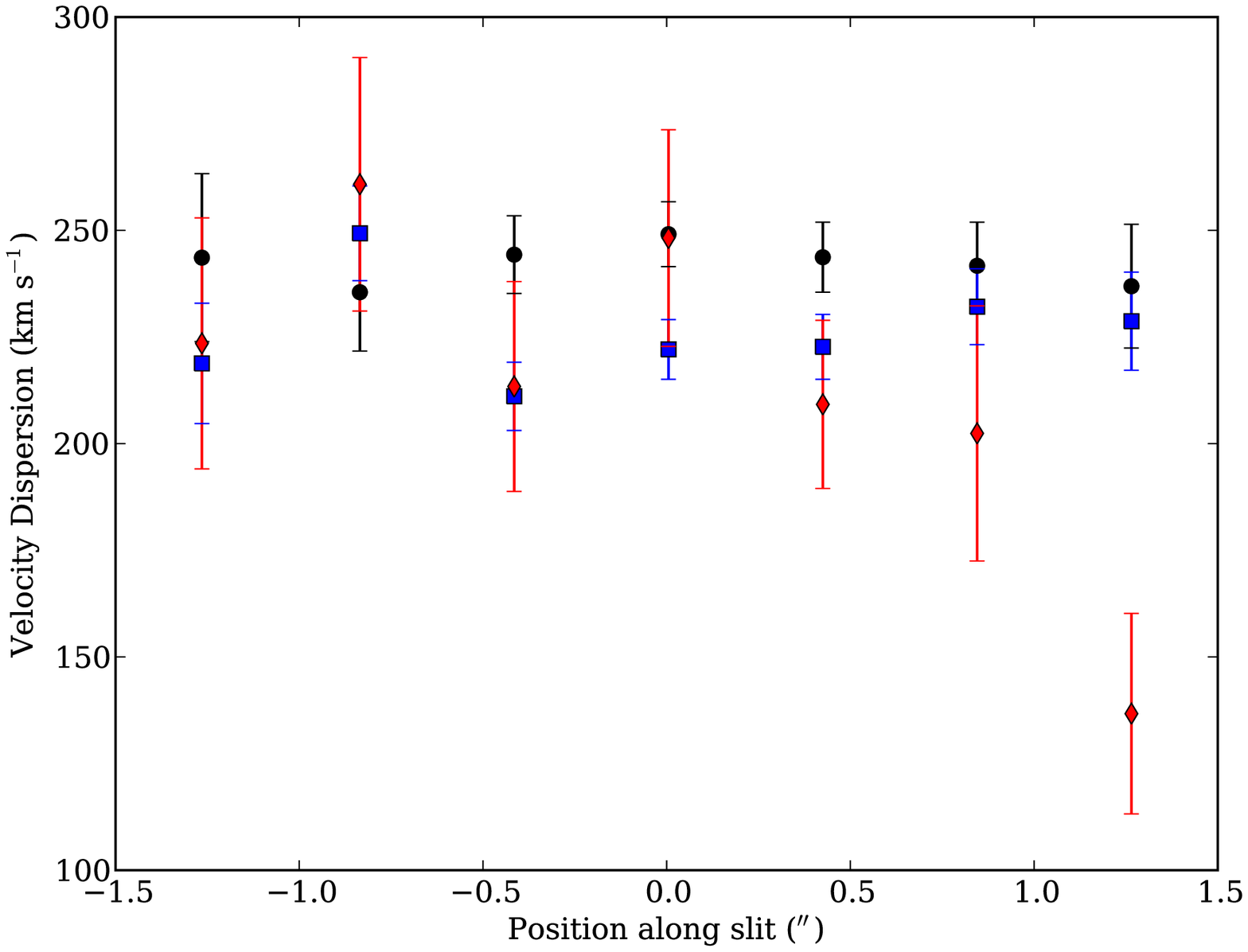}
  \end{center}
  \caption{Line-of-sight velocity (left) and velocity dispersion
    (right) profiles for the lens {\jkeck}. The black circles are for
    the central slit, the blue squares are for the slit offset
    0\farcs85 from the center, and the red diamonds are for the widest
    slit offset 2\farcs1 from the galaxy center.}
  \label{fig:velocitydata}
\end{figure*}

The stellar velocity dispersion and velocity offset was computed for
each spectrum as in \citet{Suyu2010} using the rest-frame wavelength
range 4200\,\AA\ to 5450\,\AA\ and errors were determined from Markov
chain Monte Carlo simulations. The resulting velocity and velocity
dispersion profiles for each slit position are shown in Figure
\ref{fig:velocitydata}. We compare our data with the velocity
dispersion derived from the SDSS spectrum by combining the 14 spectra
from the inner two slits and determining a composite velocity
dispersion. This is found to be 236 \kms, rather larger than the value
of 214 \kms\ found for the SDSS spectrum. We have re-analysed the SDSS
spectrum and find $\sigma = 212$ \kms, but if we mask the \ion{O}{II}
lines (as is done in the analysis of the Keck spectra) we find $\sigma
= 225$ \kms. This is still somewhat lower than, although consistent
within the errors with, the central velocity dispersion derived from
the composite Keck spectrum. However, we might expect the SDSS value
to be lower due to poor seeing pushing more flux from large radii
(where the velocity dispersion is lower) into the SDSS fibre aperture.

\section{Analysis and results}
\label{sec:analysis}

\subsection{Joint gravitational lensing and stellar dynamics analysis}
\label{ssec:code}

Here we briefly recall the general features of the {\cauldron} code,
the tool employed to carry out the combined self-consistent
gravitational lensing and stellar dynamics analysis for the galaxy in
exam, {\jkeck}. The reader is referred to BK07 for a detailed
description of the algorithm.

We characterize the lens galaxy by means of its total density
distribution $\rho(\veceta)$, where $\veceta$ is a set of parameters
describing the density profile. Via the Poisson equation, we calculate
the total gravitational potential $\Phi$ and we use it simultaneously
for both the gravitational lensing and the stellar dynamics modelling
of the data set, which typically includes the surface brightness and
velocity moments maps of the lens galaxy and the surface brightness
map of the lensed image. Both these modelling problems can be
formalized as a set of regularized linear equations, for which
standard solving techniques are readily available. Thus --- given a
combined data set --- for each choice of the parameters $\veceta$ we
can calculate the surface brightness distribution of the unlensed
source, and the weights of the elementary stellar dynamics building
blocks. In order to determine the ``best'' (in an Occam's razor sense)
density model given the data, this linear optimization scheme has been
embedded within the framework of Bayesian statistics. This allows to
objectively quantify (and therefore rank) the plausibility of each
model by means of the evidence merit function \citep[see
  e.g.][]{MacKay1999, MacKay2003}. In this way, by maximizing the
evidence, one recovers the set of non-linear parameters $\veceta$
corresponding to the best density model, i.e.\ the model which
maximizes the joint posterior probability density function (PDF),
hence called maximum \emph{a posteriori} (MAP) model.

The method as described is extremely flexible and can in principle
support any density profile, by adopting for example a completely
general pixelized density distribution. However, its current practical
implementation, the {\cauldron} algorithm, is more restricted in order
to make it computationally efficient and assumes axial symmetry
(i.e.\ a density distribution of the form $\rho(R,z)$) and a
two-integral stellar DF $f = f(E, \Lz)$, where $E$ and $\Lz$ denote
the two classical integrals of motion, i.e., respectively, energy and
angular momentum along the rotation axis. As shown in BK07, under
these assumptions it is possible to take advantage of a fast Monte
Carlo numerical implementation of the two-integral Schwarzschild
method described by \citet{Cretton1999} and
\citet{Verolme-deZeeuw2002}, which allows a dynamical model to be
built in a matter of seconds and, therefore, makes it possible to
explore large regions of the parameter space. The distinguishing
feature of this method is that the building blocks employed for the
construction of the dynamical model are constituted by two-integral
components (TICs) rather than stellar orbits as in the classical
Schwarzschild method (\citealt{Schwarzschild1979}; see
e.g.\ \citealt{Thomas2007b} and \citealt{vandenBosch2008} for modern
implementations). A TIC can be visualized as an elementary toroidal
system, completely specified by a particular choice of $E$ and
$\Lz$. TICs have simple $1/R$ radial density distributions and
analytic unprojected velocity moments, and by superposing them one can
build $f(E, \Lz)$ models for arbitrary spheroidal potentials
\citep[cf.][]{Cretton1999}: all these characteristics contribute to
make TICs particularly valuable and inexpensive building blocks when
compared to orbits.

Notwithstanding these restrictions, \citet{Barnabe2008} have shown
that {\cauldron} works robustly even when applied to simulated systems
which depart significantly from the method's assumptions (including
the assumption of axial symmetry), reliably recovering several
important global properties of such systems, in particular the slope
of the total mass density profile, which is determined within less
than 10~per cent of the correct value. When the system displays
rotation in the kinematical maps, as is the case for {\jkeck}, the
main dynamical quantities (such as the global anisotropy parameter
$\delta$, the angular momentum and the ordered to random motions
ratio, see Sect.~\ref{ssec:globdyn}) are recovered with an accuracy
of~10 to~25 per cent.

\subsection{The galaxy model}
\label{ssec:model}

Stellar dynamics \citep[e.g.][]{Gerhard2001}, strong and weak
gravitational lensing \citep[e.g.][]{Koopmans2009, Gavazzi2007} and
X-ray studies \citep[see e.g.][and references therein]{Humphrey-Buote2010} 
all concur in indicating that the total mass profile of elliptical
galaxies is remarkably well described by a single power-law model over
a large radial range.

As our model for the total mass density profile of the analyzed lens
galaxy we adopt, therefore, an axially symmetric power-law
distribution
\begin{equation}
  \label{eq:rho}
  \rho(m) = \frac{\rho_{0}}{m^{\slope}}
\end{equation}
with $\rho_{0}$ being a density scale, $0 < \slope < 3$ the
logarithmic slope of the density profile, and $m$ the elliptical
radius, i.e.\
\begin{equation}
  \label{eq:m}
  m^2 = \frac{R^2}{a_0^2} + \frac{z^2}{c_0^2} 
  = \frac{R^2}{a_0^2} + \frac{z^2}{a_0^2 q^2} ,
\end{equation}
where $c_0$ and $a_0$ are length-scales and the axial ratio $q\equiv
c_0/a_0$. The corresponding (inner) gravitational potential
$\Phi(R,z)$ associated with the distribution $\rho(m)$ can be obtained
straightforwardly by means of the classical \citet{Chandrasekhar1969}
formula, which entails the computation of a single integral. 

In case the assumption of a power-law density profile is an
oversimplified description of the true mass distribution of the
galaxy, this is expected to have visibly disrupting effects on the
reconstructed observables, in particular for the lensing ones. As
illustrated in the \citet{Barnabe2008} tests, these can include a
strongly irregular reconstructed source, with sharp transitions in
intensity between close pixels (despite the usual adoption of a
regularization term), and the presence of recognizable arc-like
features in the lens image residuals. We emphasize, however, that such
clear discrepancies have never surfaced in previous analyses of the
SLACS systems (\citealt{Czoske2008}, B09), which supports instead the
effectiveness of the simple power-law model.

With this choice, the galaxy model is therefore characterized by three
non-linear physical parameters, i.e. the slope $\slope$, the axial
ratio $q$ and the dimensionless lens strength $\talp$, which is
directly related to the normalization of the three dimensional
potential (see Appendix~B of BK07). To these, we must add the four
``geometrical'' parameters defining the configuration of the system in
the sky, i.e.\ the position angle $\PA$, the inclination $i$ and the
coordinates of the lens galaxy centre. The latter, as well as the
angle $\PA$, are typically strongly constrained by the lens image
brightness distribution and can be accurately determined by means of
fast preliminary explorations and, therefore, kept fixed afterwards in
order to decrease the number of free non-linear parameters during the
more computationally expensive optimization and error analysis
runs. If necessary, external shear can be also accounted for, by
introducing shear strength and shear angle as additional parameters.

Finally, we have three so-called hyperparameters which control the
level of the regularization in the reconstructed quantities: one for
the surface brightness distribution of the unlensed source, and two
for the TIC weights map. Their values are set by optimizing the
Bayesian posterior probability.

\subsection{Uncertainties}
\label{ssec:errors}

In order to correctly assess the model uncertainties within the
framework of Bayesian statistics, one needs to evaluate the posterior
probability distribution of the parameters, i.e., by denoting the data
set as $\vec{d}$ and the considered model or hypothesis as
$\mathcal{H}$,
\begin{equation}
  \label{eq:posterior}
  \mathcal{P}(\veceta) \equiv \pr(\veceta \, | \, \vec{d},\mathcal{H}) 
  \propto \mathcal{L}(\veceta) \times p(\veceta) \, ,
\end{equation}
where $\mathcal{L} (\veceta) \equiv \pr(\vec{d} \, | \, \veceta,
\mathcal{H})$ is the likelihood and $p (\veceta) \equiv \pr (\veceta
\, | \, \mathcal{H})$ is the prior. The individual parameter
confidence intervals can be obtained by marginalizing the posterior
over each of them.

Sampling a multidimensional distribution such as $\mathcal{P}$ is in
general a challenging and computationally expensive task. An effective
technique to tackle this problem is the nested sampling Monte Carlo
method introduced by \citet{Skilling2004} which, in calculating the
evidence, produces posterior inferences as valuable by-products. For
our error analysis, we make use of the \textsc{MultiNest} algorithm
developed by \citet{Feroz-Hobson2008} and \citet*{Feroz2009}, which
provides an efficient and robust implementation of the nested sampling
method, and has been shown to yield reliable posterior inferences even
in presence of multi-modal and degenerate multivariate distributions.

The model parameters that we consider are the ones introduced in the
previous Section, i.e. the inclination~$i$, the lens strength~$\talp$,
the slope~$\slope$ and the axial ratio~$q$ (the additional parameters
which can be estimated by means of preliminary runs, such as the lens
center and the position angle, however, are kept fixed here in order
to reduce the computational load), to which we must add the three
hyper-parameters. We formalize our ignorance by adopting flat priors
(or flat in logarithm for the hyper-parameters), constructed around
the MAP value of each parameter, and taken wide enough to include the
bulk of posterior probability distribution. Finally, we conduct a
thorough exploration of this 7-dimensional surface by launching
\textsc{MultiNest} with 400 live points\footnote{The live points
  (sometimes also called active points) are the initial samples drawn
  from the full prior distribution $p (\veceta)$, from which the
  nested sampling exploration of the posterior is started. Our choice
  of~400 live points for the relatively well-behaved 7-dimensional
  distribution at hand is very conservative. In fact, as shown by the
  test problems examined in \citet{Feroz2009}, 1000 or 2000 active
  points are sufficient for the application of \textsc{MultiNest} even
  to, respectively, highly dimensional problems (with up to 30
  parameters) or pathologically multimodal distributions (e.g. the
  egg-box likelihood presented in their Section~6.1). Moreover, if one
  is more interested in determining the marginalized posterior
  distribution of the parameters rather than in accurately calculating
  the value of the total evidence, a much smaller number of live
  points (of order~50) is shown to be already enough to obtain a
  reliable estimate.}, from which the individual marginalized
posterior probability distributions are obtained. In the following, we
quote the 99 per cent confidence interval calculated from these
distributions as our error.

\begin{table}
  \centering
  \begin{minipage}{0.80\hsize}
  \caption{Recovered model parameters for lens galaxy {\jkeck}.}
  \begin{center}
  \begin{tabular}{ c @{\hspace{4em}} c c c c}
    \hline
    \noalign{\smallskip}
    parameter & $\eta_{\textsc{map}}$ & $\eta_{-}$ & $\eta_{\mathrm{max}}$ & $\eta_{+}$ \\
    \noalign{\smallskip}
    \hline
    $ \slope $   & 2.082 & 2.055 & 2.077 & 2.119 \\
    \noalign{\smallskip}
    $ \talp $    & 0.325 & 0.317 & 0.323 & 0.331 \\
    \noalign{\smallskip}
    $ q $        & 0.602 & 0.574 & 0.603 & 0.681 \\
    \noalign{\smallskip}
    $ i $        & 68.1  & 67.8  & 68.6  & 73.6  \\
    \hline
  \end{tabular}
  \end{center}
  \label{tab:eta}

  \textit{Note.} The listed parameters are: the logarithmic slope
  $\slope$; the dimensionless lens strength $\talp$; the axial ratio
  $q$ and the inclination $i$ (in degrees) with respect to the
  line-of-sight. The second column presents the MAP parameters,
  i.e.\ the parameters that maximize the joint posterior
  distribution. Columns~3 to~5 encapsulate a description of the
  one-dimensional marginalized posterior distributions, plotted in
  Fig.~\ref{fig:errors}. The parameter $\eta_{\mathrm{max}}$
  corresponding to the maximum of that distribution is listed in
  column~4, while $\eta_{-}$ and $\eta_{+}$ are, respectively, the
  lower and upper limits of the 99 per cent confidence interval.
  \end{minipage}
\end{table}

\begin{figure*}
  \centering
  \resizebox{0.97\hsize}{!}{\includegraphics[angle=-90]
            {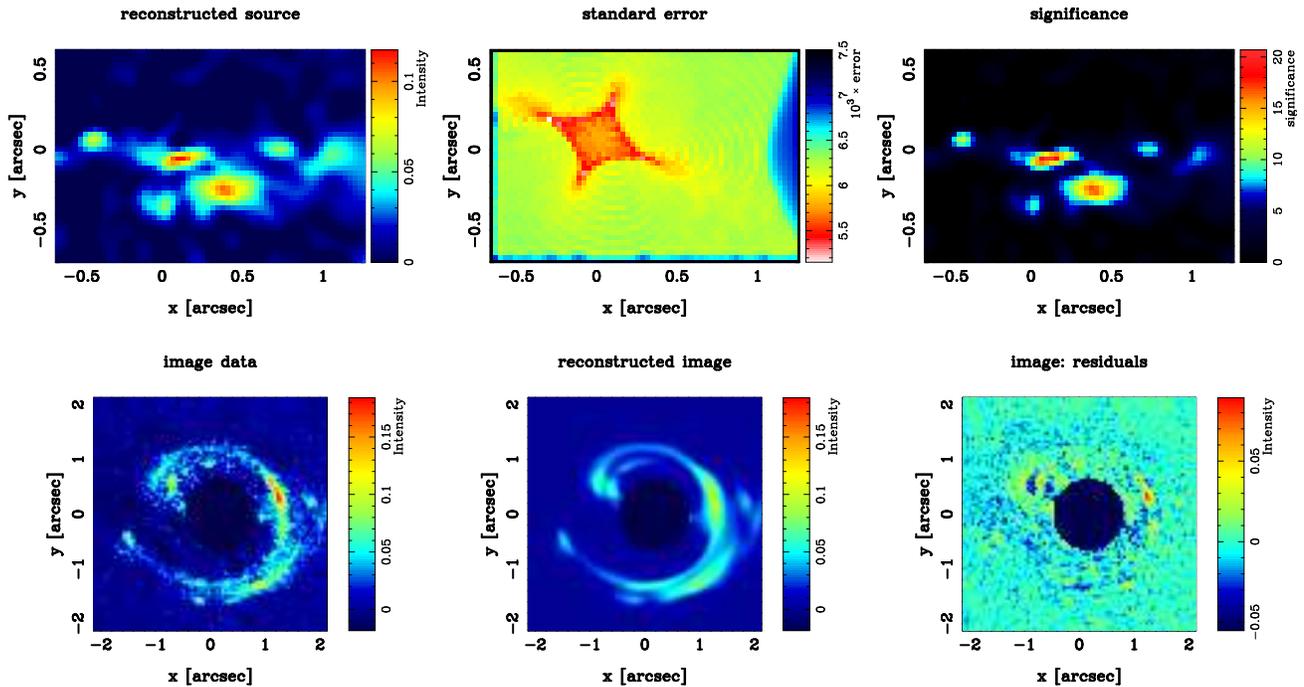}}
  \caption{Best model lens image reconstruction for the system
    {\jkeck}. \emph{Top row:} reconstructed source model; $1 \sigma$
    uncertainty on the source pixels; significance of the
    reconstructed source. \emph{Bottom row:} \textit{HST}/ACS data
    showing the lens image after subtraction of the lens galaxy; lens
    image reconstruction; residuals. In the panels, North is up and
    East is to the left.}
  \label{fig:J0728_LEN}
\end{figure*}

\begin{figure*}
  \resizebox{0.97\hsize}{!}{\includegraphics[angle=-90]
            {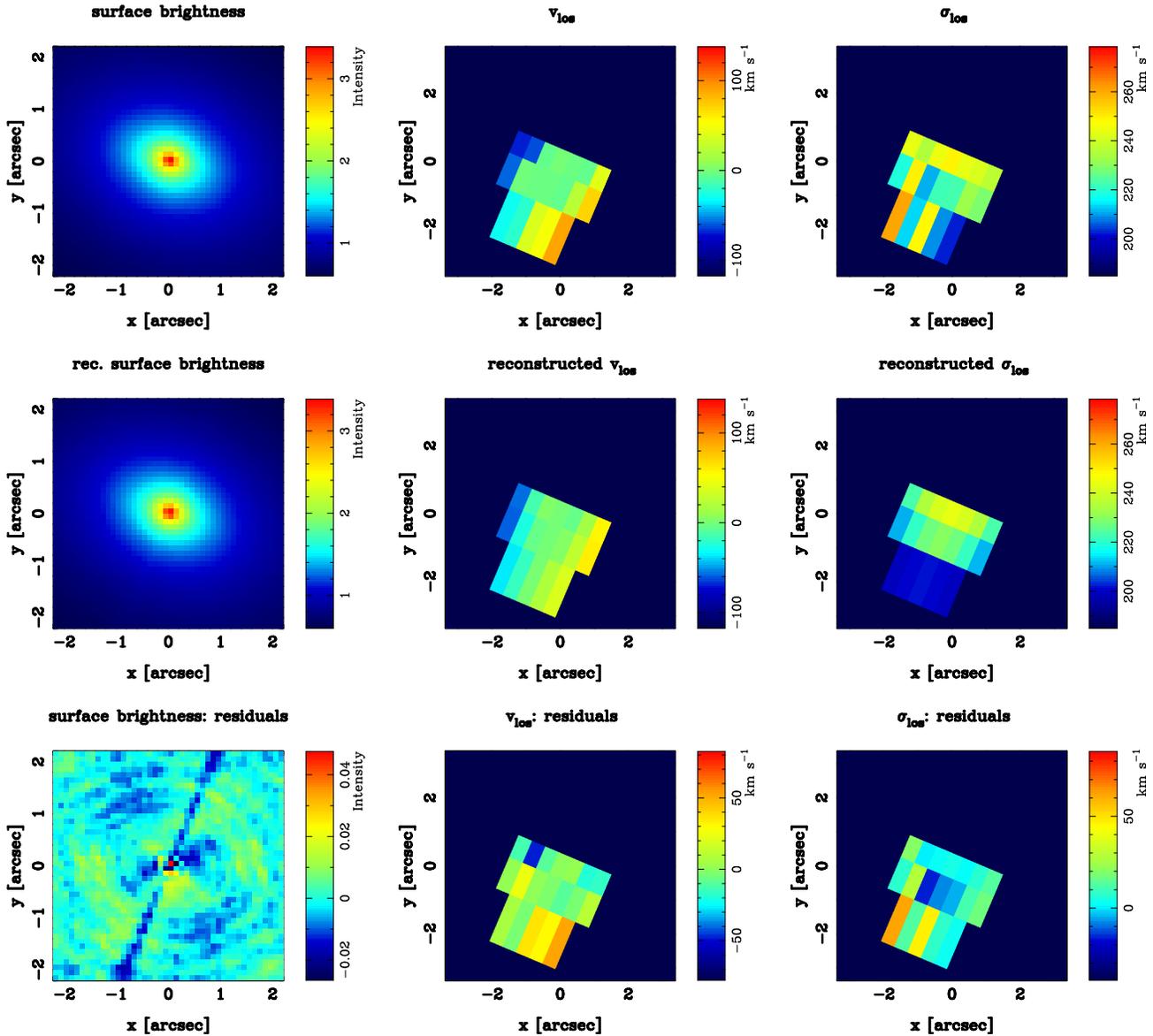}}
  \caption{Best dynamical model for the galaxy {\jkeck}. \emph{Top
      row:} observed surface brightness distribution, projected
    line-of-sight velocity and line-of-sight velocity
    dispersion. \emph{Middle row:} corresponding reconstructed
    quantities for the best model. \emph{Bottom row:} residuals. In
    the panels, North is up and East is to the left.}
  \label{fig:J0728_DYN}
\end{figure*}

\subsection{Results for the best model reconstruction}
\label{ssec:bestmodel}

The {\cauldron} code has been applied to the combined data set
described in Section~\ref{sec:observations}. In order to avoid
possible spurious effects in the lensing reconstruction, we have
masked out the central regions of the lensed image map, where the
residuals of the galaxy subtraction are still appreciable. We use the
best-fitting B-spline model of the lens galaxy as our data set for its
surface brightness distribution, to avoid ``contamination'' from the
background galaxy lensed images, which is particularly bright and
could bias the reconstructed model. A similar approach had been
followed in the analysis of lens system {\jttto} \citep{Czoske2008}.

\begin{figure*}
  \centering
  \resizebox{0.95\hsize}{!}{\includegraphics[angle=-90]
            {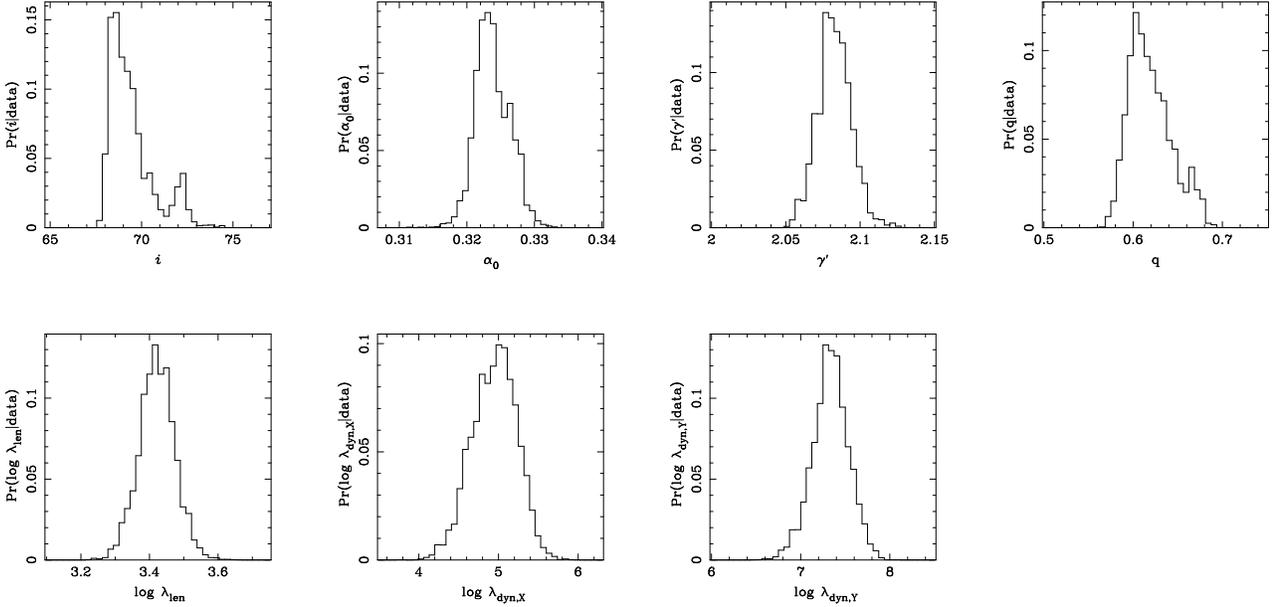}}
  \caption{Marginalized posterior probability distributions of the
    power-law model parameters (inclination, lens strength,
    logarithmic slope, and axial ratio) and hyper-parameters, obtained
    from the nested sampling exploration of the posterior. The
    uncertainties on the parameters quoted in the text are calculated
    by considering the interval around the peak which contains 99 per
    cent of the probability.}
  \label{fig:errors}
\end{figure*}

The recovered non-linear parameters for the best reconstructed model,
i.e. the maximum \emph{a posteriori} model, are presented in
Table~\ref{tab:eta}. The uncertainties on the individual parameters
are quantified by marginalizing, over each of them, the joint
posterior distribution (see Fig.~\ref{fig:errors}). The parameter
values corresponding to the maximum of the one-dimensional
marginalized posterior and the limits of the 99 per cent confidence
interval are also listed in Table~\ref{tab:eta}.

We find for {\jkeck} a logarithmic slope $\slope =
2.077^{+0.042}_{-0.022}$ (errors indicate the 99\% confidence level),
very close to the so-called isothermal (i.e.\ $\rho \sim 1/r^{2}$)
profile which appears to be a characterizing feature of early-type
galaxies, and in general agreement with previous combined lensing and
dynamics studies of the SLACS sample (\citealt{Koopmans2009},
B09). For this specific system --- by using the 3-arcsec aperture
averaged SDSS velocity dispersion measure as the only kinematic
constraint --- \citet{Koopmans2009} determine a slightly different
slope $\slope = 1.85 \pm 0.10$ (68\% CL), which, however, we find
here to be too shallow to correctly reproduce the kinematic maps. One
reason for this discrepancy is that the velocity dispersion derived
from the SDSS is lower (typically of $\sim 20$ \kms) than the value
obtained from Keck spectroscopy, as discussed in Sect.~\ref{ssec:IFU}.
Moreover, \citet{Koopmans2009} adopt a simpler dynamical model, based
on solving spherical Jeans equations, which might have difficulties in
describing a system characterized by a significant flattening like
{\jkeck}.

The recovered lens strength is $\talp = 0.323^{+0.008}_{-0.006}$. The
axial ratio of the total density distribution is found to be $q =
0.602^{+0.079}_{-0.028}$, slightly flatter than the intrinsic axial
ratio $\qstar = 0.688$ of the luminous distribution, calculated by
deprojecting the observed isophotal axial ratio $\qspro$ by making use
of the best model value $i = 68.6^{+5.0}_{-0.8}$ obtained for the
inclination. Moreover, the position angle $\PA = 67\fdg5$ is found to
be extremely close to the value inferred from the light distribution,
indicating alignment between the dark and luminous mass components in
the inner regions of the galaxy.

The marginalized posterior probability distributions of these
parameters (with the exception of the position angle, which, as
previously described, is kept fixed after the preliminary run), as
well as those of the regularization hyper-parameters, obtained as
described in Section~\ref{ssec:errors}, are shown in
Fig.~\ref{fig:errors}. We note that these represent the statistical
errors on the considered power-law model, and do not take into account
the potential systematic uncertainties due to issues in the generation
of the data sets (e.g.\ the procedure for galaxy subtraction, as
pointed out by \citealt{Marshall2007}) or to incorrect model
assumptions (as in the case of, e.g., non axially symmetric density
distribution or flattening that varies with radius). In the latter
case, a quantitative estimate of the upper limits of the systematic
errors can be obtained by looking at the findings of the
\citet{Barnabe2008} `crash-test' where the {\cauldron} code is applied
to a non-symmetric simulated galaxy. As mentioned in
Sect.~\ref{ssec:code}, the error on the logarithmic slope $\slope$ is
less than 10 per cent even in this quite extreme case. Since real
early-type galaxies are unlikely to depart from axisymmetry as
drastically as this simulated system, we expect systematic
uncertainties to remain within a few per cent level.

There is no evidence that any significant external shear effect needs
to be included in order to model this system: when the external shear
angle and strength are allowed to vary, the MAP model value of the
latter is found to be negligibly small. This is consistent with the
absence of massive structures near the lens galaxies found in
\citet{Treu2009}.  We also explored the possibility of introducing a
core radius in the density profile, without finding any improvement to
justify the inclusion of this additional parameter.

In Figure~\ref{fig:J0728_LEN} and~\ref{fig:J0728_DYN} we present ---
respectively for lensing and dynamics --- the {\jkeck} data set, the
reconstructed observables obtained for the MAP model, and the
corresponding residuals. This system displays an unusually structured
lensed image, which we find to be produced by the presence of multiple
components in the source plane. The reconstructed background source is
shown in the top-left panel of Fig.~\ref{fig:J0728_LEN}; the
top-middle and top-right panels show, respectively, the standard
errors and the significance of the reconstructed source.

The surface brightness map is reconstructed very accurately. Since the
adopted data set is a B-spline model and therefore noiseless, the
small residuals (at the 1 per cent level) are dominated by the
discreteness effects of the TIC superposition, which determines the
concentric ripples, while the structure aligned with the galaxy minor
axis is caused by the toroidal shape of the building blocks. These
undesired effects have been kept under control by increasing both the
number of TIC employed in the dynamical model (of almost a factor of
four, from $N_{E} = 10 \times N_{L_{z}} = 10$ elements adopted in the
previous studies conducted with {\cauldron} to $20 \times 9$ elements)
and the number of particles populating each TIC. An additional and
more compelling motivation for this improvement of the TIC library,
which further justifies the increased computational burden, is that it
proves to be important for the reconstruction of the kinematic
observables (particularly the velocity dispersion map) at larger
distances from the galaxy major axis, i.e. that part of our data set
which best allows us to probe the system under study beyond the
effective radius.

\subsection{Global dynamical quantities}
\label{ssec:globdyn}
\begin{figure}
  \centering
  \resizebox{0.85\hsize}{!}{\includegraphics[angle=-90]{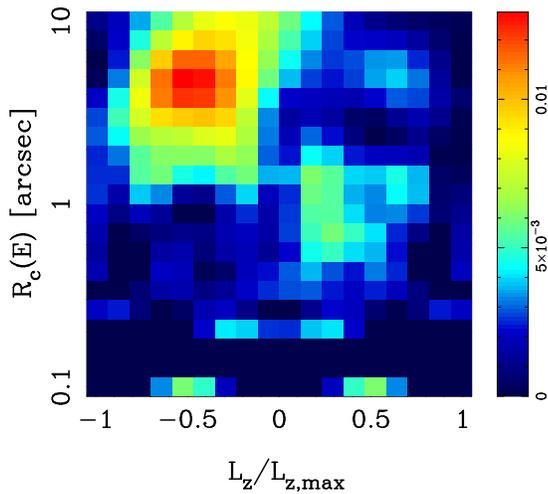}}
  \caption{Reconstruction of the weighted two-integral DF of the
    system {\jkeck} obtained from the MAP model. The value of each
    pixel in the two-integral space represents the relative
    contribution of the corresponding TIC to the stellar component of
    the modelled system.}
  \label{fig:DF}
\end{figure}

The recovered weighted stellar DF for the best combined model is
presented in Fig.~\ref{fig:DF} as a map of the relative TIC weights
over the integral space $(E, L_{z})$ (the grid in the radial
coordinate $\Rc$ is related to a grid in energy as explained in
BK07). This representation encodes, in a very compact way, much of the
information on the dynamical structure of the galaxy that can be
obtained under the assumptions of the adopted two-integral
model. However, it is often useful to distill such information into
quantities that allow for a more straightforward physical
interpretation.

The global properties of the stellar velocity dispersion tensor are
customarily encapsulated in the three anisotropy parameters
\begin{equation}
  \label{eq:AP}
  \beta \equiv 1 - \frac{\Pi_{zz}}{\Pi_{RR}}, \quad
  \gamma \equiv 1 - \frac{\Pi_{\varphi\varphi}}{\Pi_{RR}} 
  \quad \textrm{and} \quad
  \delta \equiv 1 - \frac{2 \Pi_{zz}}{\Pi_{RR} + \Pi_{\varphi\varphi}},
\end{equation}
where we indicate with
\begin{equation}
  \label{eq:AP:PI}
  \Pi_{kk} = \int \rho \sigma^{2}_{k}\, \mathrm{d}^{3} \vec{x}
\end{equation}
the total unordered kinetic energy in the coordinate direction $k$ and
$\sigma_{k}(\vec{x})$ is the velocity dispersion along the direction
$k$ \citep[see][]{Cappellari2007,BT08}. 

For {\jkeck}, we compute the integral of Eq.~(\ref{eq:AP:PI}) within a
cylinder of radius and height equal to $\Reff$, i.e. inside a region
which is very well constrained by the data, finding a mild anisotropy
$\delta = 0.08\pm0.02$, which falls within the typical
range of values for early-type galaxies both in the local Universe
\citep[][]{Cappellari2007,Thomas2009} and up to redshift $z \sim 0.35$
(see B09). Since we make use of a two-integral DF dynamical model, the
velocity dispersion tensor is isotropic in the meridional plane
(i.e. $\sigma^{2}_{R} (\vec{x}) = \sigma^{2}_{z} (\vec{x})$ for each
$\vec{x}$) and therefore $\beta \equiv 0$ by construction, and
$\gamma$ is univocally linked to $\delta$ by the relation $\gamma = 2
\delta / (\delta - 1)$. For the analyzed system, we have $\gamma =
-0.18\pm0.04$.

\begin{figure} 
  \centering
  \resizebox{0.85\hsize}{!}{\includegraphics[angle=-90]{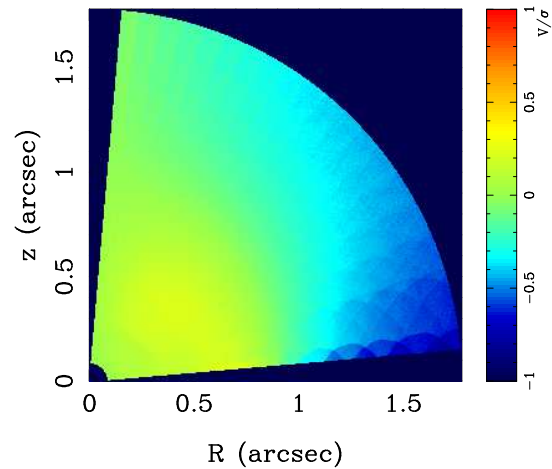}}
  \caption{Map of the local $\vphi/\bar{\sigma}$ ratio between the
    mean rotation velocity around the $z$-axis and the mean velocity
    dispersion, plotted up to $\Reff$ in the positive quadrant of
    the meridional plane}.
  \label{fig:VoS}
\end{figure}

The importance of rotation with respect to random motions is among the
most defining aspects of the dynamical structure of a stellar
system. In order to explore how this property varies with the position
in the meridional plane, we calculate the local ratio
$\vphi/\bar{\sigma}$ between the mean rotation velocity around the
$z$-axis and the mean velocity dispersion $\bar{\sigma}^{2} \equiv
(\sigma_{R}^{2} + \sigma_{\varphi}^{2} + \sigma_{z}^{2})/3$ and we
plot it in Fig.~\ref{fig:VoS} up to one effective radius. The inner
regions --- within approximately $1\arcsec$ --- are dominated by
random motions, while rotation becomes more important at large radii,
a trend somewhat reminiscent of what is found, in B09, for the fast
rotator {\jonfn} (although it should be remembered that the
$\vphi/\bar{\sigma}$ map of the latter only extends up to
$\Reff/2$). Not surprisingly, therefore, the presence of large-scale
ordered motions is reflected also in the quite high value $\jz =
0.28^{+0.05}_{-0.01}$ obtained for the intrinsic rotation parameter,
which is a dimensionless proxy for the galaxy angular
momentum (refer to B09 for the definition).

\subsection{Dark and luminous mass distribution}
\label{ssec:mass}
\begin{figure*}
  \begin{center}
    \subfigure{\label{fig:mass-chab}\includegraphics[angle=-90,width=0.48\textwidth]{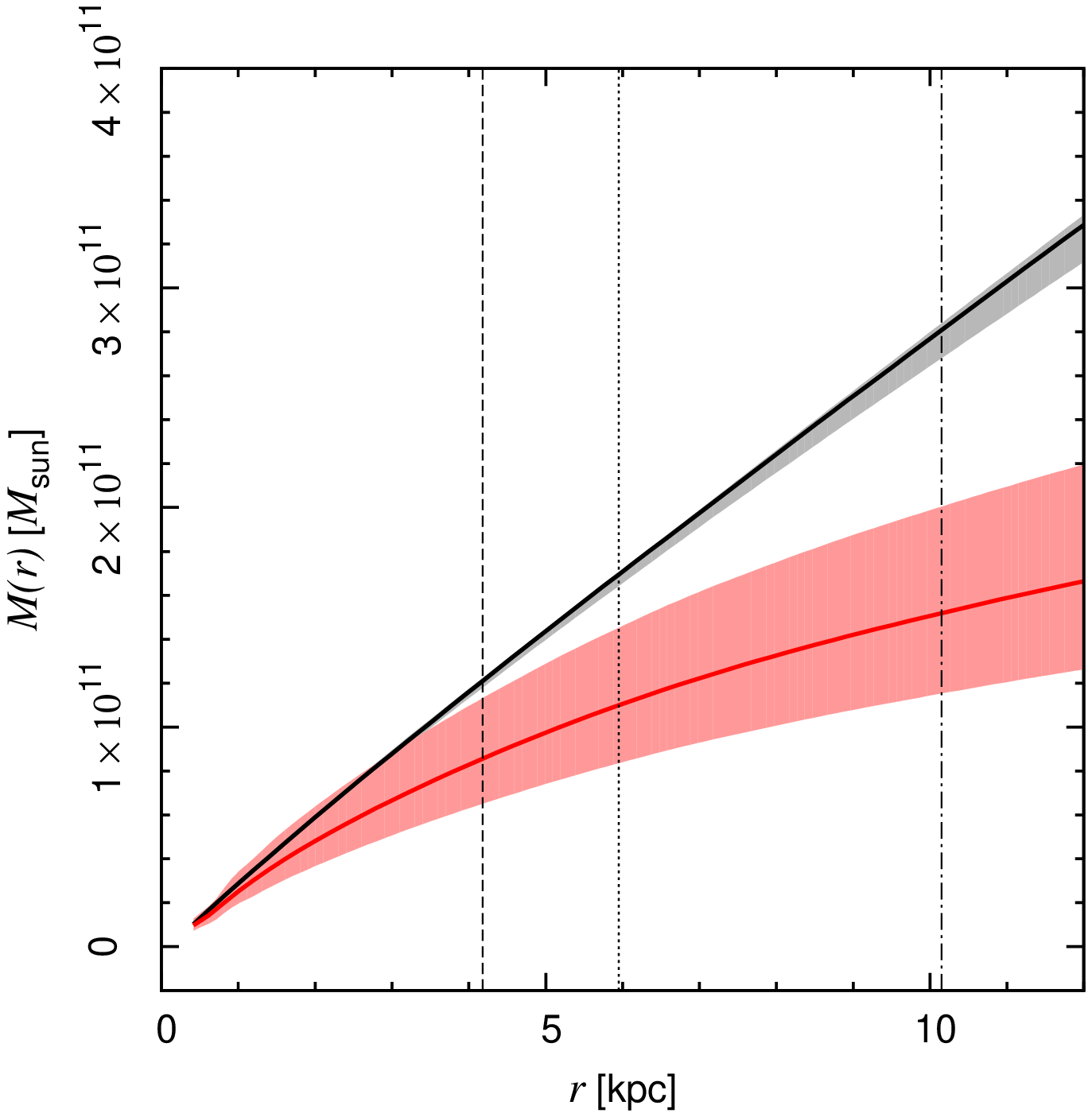}}\hfill
    \subfigure{\label{fig:mass-salp}\includegraphics[angle=-90,width=0.48\textwidth]{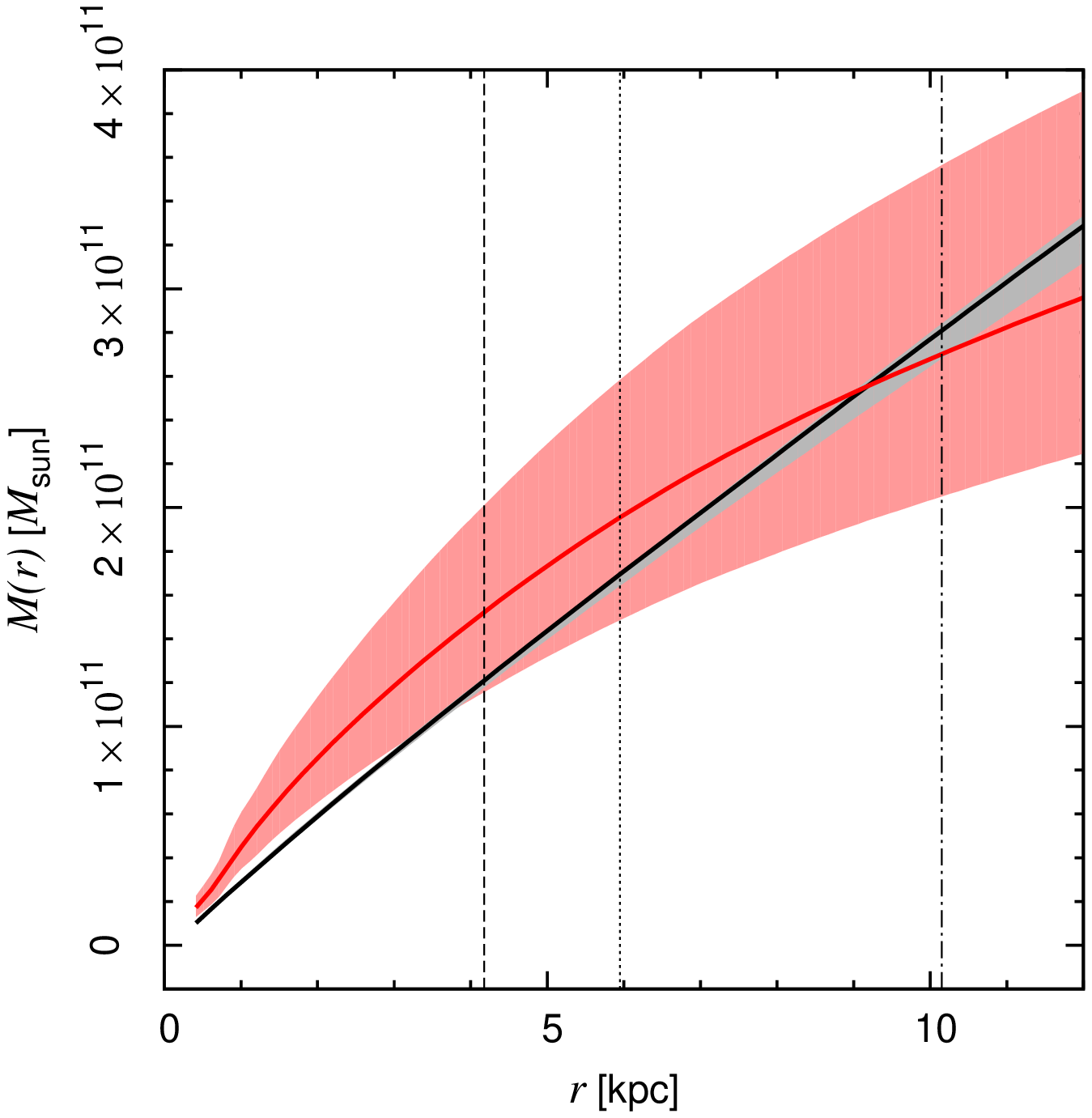}}
  \caption{Spherically averaged mass distribution for lens galaxy
    {\jkeck}. The solid black line shows the total mass profile
    obtained from the best reconstructed model, with the statistical
    uncertainty represented as a grey band. The solid red line shows
    the luminous mass profile obtained from the recovered stellar DF
    and rescaled using the stellar masses obtained from the
    \citet{Auger2009} stellar population analysis with Chabrier and
    Salpeter IMFs (left and right panel, respectively); the upper and
    lower error bars quoted in that paper set the limits for the
    red-shaded regions. The vertical lines provide an indication of
    the region probed by the data set, showing the three-dimensional
    radius $r$ which equals the Einstein radius (dashed line), the
    effective radius (dotted line) and the outermost boundary of the
    kinematic maps (dash-dotted line).}
  \end{center}
  \label{fig:massprof}
\end{figure*}

The spherically averaged profile of the galaxy total mass
corresponding to the best reconstructed model (solid black curve in
Fig.~\ref{fig:massprof}) can be calculated straightforwardly from
Eq.~(\ref{eq:rho}), while the radial profile of the luminous component
is obtained from the recovered stellar DF. However, since within our
method stars are treated as tracers of the total potential, the
normalization of the luminous profile is not fixed, and must be
constrained by means of an independent determination of the stellar
mass-to-light ratio or by introducing additional assumptions.

One particularly informative assumption, known as the ``maximum
bulge'' approach, consists in maximally rescaling the luminous mass
distribution without exceeding the total mass profile at any
radius. This provides a consistent and robust way to assess a lower
limit for the dark matter fraction in the analyzed system (under the
hypothesis that the stellar mass-to-light ratio does not change too
drastically with the position in the galaxy). By applying this
procedure to {\jkeck}, we determine a value $\log(M_{\star}/M_{\sun})
= 11.50$ for the total stellar mass. With this normalization of the
luminous mass profile, one finds a dark matter fraction of 16 per cent
of the total mass within the (three-dimensional) spherical radius $r =
\Reff/2$, which rises to 28 per cent at $r = \Reff$ --- corresponding
to about $6$ kpc --- and up to almost 40 per cent at $r \sim 10$ kpc,
which is approximately the outer limit of the region over which we
have direct information from the stellar kinematic maps. This result
is consistent with the findings of purely dynamical studies of the
inner regions of early-type galaxies in the local Universe
\citep[e.g.][]{Gerhard2001,Cappellari2006,Thomas2007b}. It is also in
good agreement with the conclusions of our previous analysis of six
SLACS lens systems (where the same maximum bulge prescription was
adopted), for which, however, the kinematic data set does not extend
beyond $\Reff$, with the exception of galaxy {\jonfn} (see B09). The
corresponding upper limit for the stellar mass-to-light ratio in the
$B$ band of {\jkeck} is $M_{\star}/L_{B} = 3.14$, which is in the
lower end of the typical values reported for slow-rotating elliptical
galaxies in the local Universe \citep{Kronawitter2000, Gerhard2001,
  Trujillo2004}.  This is not too surprising, since the
$M_{\star}/L_{B}$ is expected to increase by a significant amount (a
factor of $1.4$ according to \citeauthor{Treu2002}
\citeyear{Treu2002}) between $z = 0.2$ and $z = 0$, simply due to
passive evolution of stellar populations. Moreover, the evidence of
rotation at large radii (cf.\ \S~\ref{ssec:globdyn}) might indicate
the presence of a disk component, typically characterized by a lower
mass-to-light ratio.

It is interesting to compare this maximum bulge upper limit with the
stellar masses determined from stellar population analysis. By
applying a novel Bayesian stellar population analysis code to
multi-band imaging data, \citet{Auger2009} determine for galaxy
{\jkeck} --- without including any priors from lensing --- a value
$\log(M_{\star}/M_{\sun}) = 11.44 \pm 0.12$ for a \citet{Chabrier2003}
IMF and a value $\log(M_{\star}/M_{\sun}) = 11.69 \pm 0.12$ for a
\citet{Salpeter1955} IMF (quoted errors are 1-sigma). We note that,
since the thermally pulsing asymptotic giant branch stars do not
dominate the luminosity of old stellar populations, the
\citet{Bruzual-Charlot2003} models used by \citet{Auger2009} should
not be biased by ignoring them. In fact, the stellar masses for SLACS
galaxies are found to be consistent with the masses determined from
\citet{Maraston2005} models \citep[see][]{Treu2010}.

By using these stellar mass values to rescale the (spherically
averaged) luminous profile, one obtains the red curves shown in
Fig.~\ref{fig:massprof}. It can be clearly seen that the Salpeter IMF
produces a stellar mass distribution which (unphysically) overshoots
the total mass profile up to and beyond the effective radius. The
Chabrier IMF, on the other hand, yields a physically acceptable
luminous profile, which, interestingly, is also very close to the one
determined through the maximum bulge assumption.

\citet{Treu2010} argue -- based on the mass determinations from
lensing, dynamics and stellar populations synthesis models for 56
SLACS systems -- that early-type galaxies cannot have both universal
IMF and dark matter profiles. In fact, if a universal Navarro, Frenk
\& White (NFW, \citealt*{Navarro1996}, \citeyear{Navarro1997}) halo is
assumed, the IMF shows a trend with velocity dispersion: a
Chabrier-like normalization is more appropriate for less massive
systems with $\sigma$ of the order of $200$ km s$^{-1}$, while more
massive galaxies are better described with a Salpeter-like IMF. The
results of our detailed analysis of {\jkeck} are consistent with this
general picture. With a SDSS velocity dispersion $\sigma = 214 \pm 11$
km s$^{-1}$, this system definitely belongs to the lower mass end of
the SLACS sample, and its halo profile is consistent with NFW (as
discussed in the next Section). Clearly, since this conclusion is
based on only one system, a full analysis of the sample still needs to
be done to further strengthen this tentative trend.

\subsection{Is the isothermal profile consistent with a NFW halo?}

The combined analysis of {\jkeck} has provided us with the total mass
density profile of this galaxy within its inner regions. As discussed
in the previous Section, and visualized in Fig.~\ref{fig:massprof},
the luminous component alone cannot account for this mass profile over
the entire radial range, even if its contribution is maximized (unless
the stellar mass-to-light ratio changes with radius in a very
fine-tuned manner). One, therefore, needs to invoke an additional mass
component characterized by the specific profile that complements the
luminous distribution. Interestingly, we find that the mass
distribution of this dark component can be consistent with a NFW halo
profile.

To show this, we attempt to describe the (spherically-averaged) total
mass distribution of the galaxy as the sum of the luminous component,
calculated as before from the recovered stellar DF, and a dark
component modelled as a standard NFW halo, i.e.
\begin{equation}
  \label{eq:rho_NFW}
  \rho_{\mathrm{NFW}}(r) = \frac{\rho_{\mathrm{crit}} \delta_{c}}
                           {(r/\rb)\,(1 + r/\rb)^{2}} \, ,
\end{equation}
where $\rho_{\mathrm{crit}}$ is the critical density, the
characteristic overdensity of the halo $\delta_{c}$ is a dimensionless
parameter connected to the halo concentration $c$ by the relation
\begin{equation}
  \label{eq:deltac}
  \delta_{c} = \frac{200}{3} \frac{c^{3}}
               {\left[ \ln (1+c) - c/(1+c) \right]}
\end{equation}
and $\rb$ denotes the break radius. We adopt $\rb = 25$ kpc, based on
the \citet{Gavazzi2007} weak lensing analysis of the SLACS sample and
the consideration that {\jkeck} is slightly less massive than the
average SLACS system (for which $\langle \sigma_{\mathrm{SDSS}}
\rangle \simeq 250$ km s$^{-1}$).

\begin{figure}
  \centering
  \resizebox{1.00\hsize}{!}{\includegraphics[angle=-90]
            {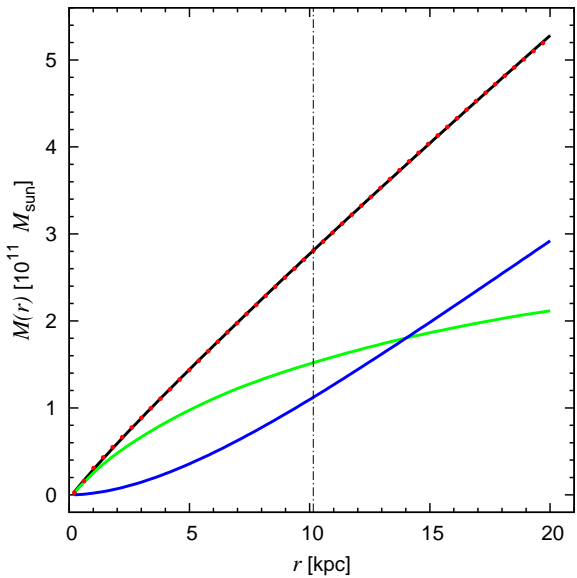}}
  \caption{Best-fit decomposition of the total mass profile of
    {\jkeck} (black curve) in the sum of a NFW dark matter halo with
    break radius $\rb = 25$ kpc (blue curve) and the luminous profile
    obtained from the recovered stellar DF (green curve). The dotted
    red line which very closely matches the black curve over the
    whole radial interval is the profile obtained by adding up the
    luminous and dark components. The vertical dash-dotted line
    indicates the outermost boundary of the kinematic maps.}
  \label{fig:NFW}
\end{figure}

We then determine the (non-negative) normalization coefficients for
the dark and luminous distribution which allow to best reproduce (in a
least-square sense) the total mass profile of the galaxy. Remarkably
--- despite the fact that the only two free parameters here are the
rescaling factors --- the superposition of these two simple components
proves to be enough to reproduce the total profile with great accuracy
over the whole radial range covered by the observations, as shown in
Fig.~\ref{fig:NFW}. Moreover, the combined profile remains consistent
with the one predicted by a nearly isothermal density distribution
even beyond $10$ kpc and up to the break radius.

The luminous profile obtained in this way almost coincides with the
one determined by means of the maximum bulge approach, with
$\log(M_{\star}/M_{\sun}) = 11.50 $, while the normalization for the
dark halo profile translates into a concentration parameter $c \simeq
11$, from which one infers a virial radius $r_{200} = c \rb \simeq
280$ kpc and a halo mass $M_{200} \simeq 3.1 \times 10^{12}
M_{\sun}$. This is a mildly high value for the concentration when
compared with the range $c \sim 3 - 10$ obtained from numerical
simulations of relaxed dark matter haloes of corresponding mass
\citep*[see in particular][]{Maccio2008}.  However, concentrations
higher than the theoretical predictions are found in dynamical studies
of slow-rotating early-type galaxies \citep[see e.g.][and references
  therein]{Romanowsky2010}. We note that, in our case, lower
concentrations are obtained by setting a larger value for the break
radius, e.g. $c \sim 10$ for $\rb = 30$ kpc, while the fit to the
total mass profile becomes only slightly worse, with discrepancy of a
few percent.


\section{Conclusions}
\label{sec:conclusions}

We have carried out a detailed study of the mass profile and dynamical
structure of the inner regions of the early-type lens galaxy {\jkeck},
located at a redshift $z = 0.21$, using a composite data set
consisting of \emph{HST}/ACS high-resolution imaging and
two-dimensional kinematic maps constructed from long-slit
spectroscopic observations obtained with the Keck instrument LRIS (the
slit --- aligned with the galaxy major axis --- has been positioned at
three different heights along the minor axis, allowing to mimic
integral-field spectroscopy). We have modelled the system by making
use of the {\cauldron} code for combined gravitational lensing and
stellar dynamics analysis, which operates under the assumptions of
axial symmetry and two-integral stellar DF. With respect to sample
studies of the SLACS lens galaxies such as \citet{Koopmans2009}, the
approach used here presents a number of improvements: it employs a
self-consistent framework where the same total potential is used for
both lensing and dynamics; it allows one to construct genuine
axisymmetric dynamical models (albeit restricted to two integrals of
motions); it extracts much more information from the data set, making
use --- in addition to the lensed image --- of the lens galaxy surface
brightness and velocity moments maps rather than being limited to a
single measure of SDSS velocity dispersion. This permits a much
greater level of detail to be recovered when modelling the system.

{\jkeck} is the first galaxy for which it has been possible to conduct
this kind of in-depth combined analysis by taking advantage of a
kinematic data set that extends well past the effective radius: the
outermost pixels of the velocity moments maps probe a region up to a
distance of about $3$ arcsec from the center, corresponding to $\sim
1.7$ $\Reff$. For comparison, in the sample of six SLACS lens galaxies
examined in B09, the outermost boundary $\Rkin$ of the kinematic maps
(obtained with VLT VIMOS integral-field spectroscopy) is in the range
$0.30 - 0.85$ $\Reff$, with the single exception of system {\jonfn}
for which $\Rkin$ exceeds the effective radius of about $15$ per cent.

The main results of the analysis are summarized and discussed below:
\begin{enumerate}
\item The total mass density profile of the galaxy inner regions, up
  to a radius of $\sim$ 1.7 $\Reff$, is found to be satisfactorily
  described by a simple axisymmetric distribution $\rho \propto
  1/m^{\slope}$, with a logarithmic slope $\slope =
  2.08^{+0.04}_{-0.02}$ (99~\% CL). This is very close to the
  isothermal profile (i.e.\ $\rho \propto 1/m^{2}$), which seems to be
  an ubiquitous characteristic of early-type galaxies at least up to
  redshift $z \sim 1$, as revealed by a number of dynamics, X-ray and
  combined lensing and dynamics studies (see
  e.g. \citealt{Kronawitter2000}, \citealt{Gerhard2001},
  \citealt{Humphrey-Buote2010}, \citealt{Koopmans2009} and references
  therein).  Moreover, weak lensing studies show that the total
  density profile remains consistent with being isothermal also well
  beyond the galaxy inner regions, up to radial distances of about 100
  effective radii \citep{Gavazzi2007}.

  The physical mechanisms which cause the total density distribution
  to be shaped into this particular structure, with little but non
  negligible intrinsic scattering ($\lesssim 10$ per cent in $\slope$,
  see B09 and \citealt{Koopmans2009}), are not well understood. The
  apparent ``conspiracy'' between the luminous and dark components to
  generate a nearly isothermal combined profile appears even more
  surprising when considering that numerical simulations including gas
  dynamics find dark matter density profiles which depend both on the
  amount of baryons and on the details of the assembly processes
  \citep[see e.g.][]{Tissera2009,Abadi2009}. 

  Interestingly, despite the predicted complications, for {\jkeck} we
  find that it is possible to reproduce very accurately the
  (spherically averaged) total mass distribution by combining two very
  intuitive and simple building blocks: (1)~the luminous mass profile
  obtained from the stellar DF, almost maximally rescaled and (2)~a
  NFW dark matter halo (with a break radius $\rb = 25$ kpc and a
  concentration parameter $c \sim 11$).

\item We find the total density distribution to be quite flattened
  within the probed region, with an axial ratio $q =
  0.60^{+0.08}_{-0.03}$, which is flatter than the axial ratio
  $\qstar$ of the luminous distribution (obtained by using the best
  model recovered inclination to deproject the two-dimensional
  isophotal axial ratio). This characteristic differentiates {\jkeck}
  from the six lens galaxies studied in B09, for which $q/\qstar
  \gtrsim 1$.

\item The system is characterized by a very mild anisotropy $\delta =
  0.08\pm0.02$. On examining the dynamical structure of the galaxy by
  means of the $\vphi/\bar{\sigma}$ map, one notices that the
  contribution of ordered motions becomes more important outside the
  inner regions, which determines the moderately high value $\jz =
  0.28$ for the dimensionless angular momentum parameter. This result
  is obtained by integrating within a cylindrical region of radius and
  height equal to $\Reff$. If the integration is limited to $\Reff/2$,
  in order to allow a direct comparison with results of B09, one finds
  a lower $\jz = 0.18$, fully consistent with the typical values
  obtained for the galaxies in that sample (with the exception of the
  clearly fast-rotating {\jonfn}).

\item Under the assumptions of maximum bulge and position-independent
  stellar mass-to-light ratio, we determine for the dark matter
  fraction a lower limit of 28 per cent within the spherical radius $r
  = \Reff$. Within $r \sim 10$ kpc, i.e.\ the approximate extension of
  the area directly probed by the kinematic data, the contribution of
  the dark matter to the total cumulative mass is about 40 per cent,
  almost matching the luminous component. This is in agreement with
  the findings of dynamical studies of nearby ellipticals
  \citep{Gerhard2001,Cappellari2006,Thomas2007} as well as with the
  combined lensing and dynamics analysis of six SLACS galaxies at $z =
  0.08 - 0.33$ (\citealt{Czoske2008}, B09). Interestingly, numerical
  simulations of early-type galaxy formation from cosmological initial
  conditions also predict a dark matter fraction of about $20 - 40$
  per cent within this radius \citep{Naab2007}.

  In alternative to the previous approach, we have also rescaled the
  luminous profile by using the stellar masses calculated from the
  \citet{Auger2009} stellar population analysis. The obtained luminous
  mass distribution is too high in the case of a Salpeter IMF,
  exceeding the total mass in places, while for a Chabrier IMF it
  remains lower than $M_{\mathrm{tot}}(r)$ and, moreover, close to the
  profile predicted under the maximum bulge hypothesis. This suggests,
  in agreement with the conclusions of \citet{Treu2010}, that a
  Chabrier functional form might be more suited to describe the IMF
  for less massive early-type galaxies such as {\jkeck}.

  In order to further test if this description is correct, we plan to
  extend this study to a wider sample of SLACS lens galaxies, covering
  a broad range of velocity dispersions $\sigma \approx 200 - 350$ km
  s$^{-1}$, for which two-dimensional kinematic data sets are
  available.

\end{enumerate}

In conclusion, despite being located at a redshift greater than $0.2$,
the system {\jkeck} shows structural characteristics --- namely nearly
isothermal total density profile, dark matter fraction, anisotropy
parameter $\delta$, local ratio of ordered to random motions ---
broadly consistent with what is observed in the nearby Universe for
early-type galaxies of comparable luminosity and velocity dispersion
(e.g.\ \citealt{Gerhard2001}, \citealt{Thomas2007},
\citealt{Cappellari2007}). The upper limit for the B-band stellar
mass-to-light ratio, obtained from the maximum bulge assumption, is
also in line with the values determined for local ellipticals
\citep[e.g.][]{Kronawitter2000}, once the ageing of the stellar
populations is taken into account. This study, therefore, provides an
indication that the density profile as well as the global dynamical
structure of the inner regions of massive ellipticals did not undergo
any dramatic change or significant evolution across the last two
billion years.


\section*{Acknowledgments}

M.B. is grateful to Phil Marshall and Simona Vegetti for useful
discussion and to Farhan Feroz for his help with
\textsc{MultiNest}. M.B. acknowledges support from the Department of
Energy contract DE-AC02-76SF00515. T.T. acknowledges support from the
NSF through CAREER award NSF-0642621, and from the Packard Foundation
through a Packard Fellowship. L.K. is supported through an NWO-VIDI
program subsidy (project number 639.042.505). Some of the data
presented herein were obtained at the W.M. Keck Observatory, which is
operated as a scientific partnership among the California Institute of
Technology, the University of California and the National Aeronautics
and Space Administration. The Observatory was made possible by the
generous financial support of the W.M.~Keck Foundation. The authors
wish to recognize and acknowledge the very significant cultural role
and reverence that the summit of Mauna Kea has always had within the
indigenous Hawaiian community.  We are most fortunate to have the
opportunity to conduct observations from this mountain. This paper is
also based on observations made with the NASA/ESA Hubble Space
Telescope, obtained from the data archive at the Space Telescope
Institute. STScI is operated by the association of Universities for
Research in Astronomy, Inc. under the NASA contract NAS 5-26555. This
work was supported by NASA through HST grants GO-10886 and 11202.


\bibliography{my_bibliography}

\label{lastpage}

\clearpage

\end{document}

%% file: definitions.tex
\DeclareRobustCommand{\ion}[2]{%
  \textup{#1\,{\mdseries\textsc{#2}}}%
}

\renewcommand\vec[1]{\bmath{#1}}

\newcommand{\Rc}{R_{\rm c}}

\newcommand{\PA}{\vartheta_{\mathrm{PA}}}
\newcommand{\talp}{\alpha_{0}}
\newcommand{\Lz}{L_{z}}

\newcommand{\veceta}{\vec{\eta}}

\newcommand{\cauldron}{\textsc{cauldron}}

\newcommand{\jonfn}{SDSS\,J0959}

\newcommand{\jttto}{SDSS\,J2321}
\newcommand{\jkeck}{SDSS\,J0728}

\newcommand{\slope}{\gamma'}

\newcommand{\qstar}{q_{\star}}
\newcommand{\qspro}{q_{\mathrm{\star, 2D}}}
\newcommand{\Reff}{R_{\mathrm{e}}}

\newcommand{\REin}{R_{\mathrm{Einst}}}

\newcommand{\Rkin}{R_{\mathrm{kin}}}

\newcommand{\jz}{j_{z}}

\newcommand{\vphi}{\langle v_{\varphi} \rangle}

\newcommand{\rb}{r_{\mathrm{b}}}
\newcommand{\pr}{\mathrm{Pr}}
\newcommand{\kms}{km s$^{-1}$}